\documentclass[pra,twocolumn,showpacs,amsmath,amssymb]{revtex4}

\usepackage{graphicx}
\usepackage{dcolumn}
\usepackage{bm}
\usepackage{color}

\begin{document}

\preprint{LA-UR 11-03191; Draft Version 12.0 December 18, 2011}

\title{Effect of an electric field on superfluid helium scintillation 
  produced by $\alpha$-particle sources}

\author{T.~M.~Ito }
\email[Corresponding author. Electronic address: ]{ito@lanl.gov}
\author{S.~M.~Clayton}
\author{J.~Ramsey}
\affiliation{Los Alamos National Laboratory, Los Alamos, NM 87545}

\author{M.~Karcz}
\author{C.-Y.~Liu}
\author{J.~C.~Long}
\author{T.~G.~Reddy}
\affiliation{Department of Physics Indiana University, Bloomington, IN 47405}

\author{G.~M.~Seidel}
\affiliation{Department of Physics, Brown University, Providence, RI, 02912}

\date{\today}

\begin{abstract}
We report a study of the intensity and time dependence of
scintillation produced by weak $\alpha$-particle sources in superfluid
helium in the presence of an electric field ($0-45$~kV/cm) in the
temperature range of 0.2~K to 1.1~K at the saturated vapor
pressure. Both the prompt and the delayed components of the
scintillation exhibit a reduction in intensity with the application of
an electric field. The reduction in the intensity of the prompt
component is well approximated by a linear dependence on the electric
field strength with a reduction of 15\% at 45~kV/cm. When analyzed
using the Kramers theory of columnar recombination, this electric
field dependence leads to the conclusion that roughly 40\% of the
scintillation results from species formed from atoms originally
promoted to excited states and 60\% from excimers created by
ionization and subsequent recombination with the charges initially
having a cylindrical Gaussian distribution about the $\alpha$ track of
60~nm radius.  The intensity of the delayed component of the
scintillation has a stronger dependence on the electric field strength
and on temperature.  The implications of these data on the mechanisms
affecting scintillation in liquid helium are discussed.
\end{abstract}

\pacs{34.50.Gb, 33.50.-j,82.20.Pm}

\maketitle

\section{Introduction}
The phenomenon of liquid helium (LHe) scintillation due to passage of
charged particles was discovered in the late
1950's~\cite{THO59,FLE59}. Since then rather extensive studies have
been conducted, motivated both by its intrinsic interest (including an
interest in illuminating the behavior of ions and neutrals in
superfluid helium) and by the application of liquid helium as a
particle detector~\cite{BAN96,MCK03}. (For a brief review of early
work, see Ref.~\cite{DUN71}. A more recent review can be found in the
introduction of Ref.~\cite{MCK03}.)
           
The following picture has emerged from these studies as the process
for liquid helium scintillation production (see
e.g. Refs.~\cite{MCK03,MCK02,ADA01,GUO09}): when a charged particle
passes through liquid helium, it deposits energy to the medium, part
of which goes to ionization, creating electron-ion pairs along its
track. The deposited energy also goes to promoting atoms to excited
states. The $W$ value, the average energy required to produce one
electron-ion pair, in liquid helium is $W\sim
43$~eV~\cite{JES55}. Electrons and ions then thermalize with the
liquid helium. The electron subsequently forms a ``bubble'' in the
liquid, pushing away surrounding helium atoms as a consequence of
Pauli exclusion.  The He$^+$ ion, on the other hand, forms a ``helium
snowball'' attracting surrounding helium atoms. The bubbles and
snowballs then recombine quickly ($\sim 10^{-10}$~s) forming excited
helium molecules (excimers). The molecules are formed in the triplet
and singlet states. The lowest-energy singlet state molecule,
He$_2$(A$^1\Sigma_u^+$), radiatively decays in less than 10$^{-8}$ s
to the (unbound) ground state, emitting a $\sim 16$~eV
extreme-ultraviolet (EUV) photon, contributing to the prompt component
of scintillation. Excited atoms in singlet states also contribute to
the prompt scintillation signal. The triplet state molecule,
He$_2$(a$^3\Sigma_u^+$), on the other hand, has a lifetime of
$\sim 13$~s in liquid helium~\cite{MCK99,KON91}. In a high excitation
density environment, however, the triplet state excimers can also be
destroyed through the Penning ionization process
\begin{equation}
\label{eq:penning1}
{\rm He}_2^* + {\rm He}_2^* \rightarrow 3{\rm He} + {\rm He}^+ + e^-,
\end{equation}
or
\begin{equation}
\label{eq:penning2}
{\rm He}_2^* + {\rm He}_2^* \rightarrow 2{\rm He} + {\rm He}_2^{\;+} + e^-.
\end{equation}
If a singlet excimer is formed as a result of Penning ionization after 
the first $10^{-7}$ s, then it
makes a contribution to the delayed scintillation component.

\begin{figure}
\includegraphics[width=9cm]{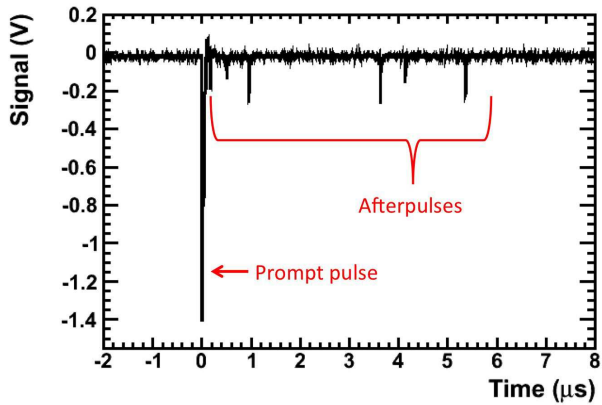}
\caption{(Color online) Oscilloscope trace of the PMT anode signal of a typical
  event. The variation in the afterpulse amplitude is due to the
  spread in the amplitude of single PE pulses, which is about 30\% for
the PMT employed in this experiment.}
\label{fig:PMTsignal}
\end{figure}

The study of liquid helium scintillation has been conducted using
radioactive sources, such as $\beta$, conversion electron, and
$\alpha$ sources as well as electron beams. There were also
experiments that studied liquid helium scintillation induced by the
products of the $^3$He($n$,$p$)$^3$H reaction~\cite{MCK03,ARC06}. Various
characteristics of the scintillation light depend on the type of
ionizing particles employed. When a $\beta$ or conversion electron
source is used, the scintillation light yield is known not to show a
strong temperature dependence, whereas when the scintillation is
induced by $\alpha$ particles, the scintillation yield does exhibit a
temperature dependence. In addition, a difference in time dependence
of the scintillation signal has been observed between $\alpha$-induced
scintillation and $\beta$-induced scintillation~\cite{MCK03}:
$\alpha$-induced scintillation light shows a long lived component that
decays with a $1/t$ dependence following the prompt pulse. This slow
component shows up as a series of single photoelectron (PE) pulses
following the prompt pulse in an experiment in which the EUV
scintillation is wavelength-shifted and detected using a
photomultiplier (PMT). (Thus these single PE pulses are called
``afterpulses''.) A typical PMT oscilloscope trace from our experiment
described in this paper is shown in Fig.~\ref{fig:PMTsignal}.

One distinct difference between $\alpha$ tracks and electron tracks is
the ionization density. For example, a 5.5-MeV $\alpha$ particle
deposits energy into superfluid liquid helium ($\rho =
0.145$~g/cm$^3$) at a rate of $2.0\times 10^4$~eV/$\mu$m on average
along its 0.27~mm long track\footnote{The maximum energy deposition
  rate is $3.9\times 10^4$~eV/$\mu$m and occurs when the remaining
  kinetic energy is 0.65~MeV.}, whereas for a 500~keV electron, the
average energy deposition rate is 40~eV/$\mu$m~\cite{NISTSTAR}. With
$W=43$~eV, it follows that ionization events are separated by an
average distance of $\sim 2$~nm and $\sim 900$~nm in $\alpha$ and
electron tracks, respectively. With an average separation of several
tens of nanometers between the electron and its parent
ion~\cite{TEN63}, ionization events in $\alpha$ tracks overlap with
each other whereas electron-ion pairs are well separated in electron
tracks. As we will discuss later, many of the differences in the
characteristics between $\alpha$-induced scintillation and
electron-induced scintillation can be attributed to the difference in
the ionization density. Although the amount of data on LHe
scintillation induced by protons and tritons from the $^3$He($n$,$p$)$^3$H 
reaction is rather limited, it shows qualitatively similar
characteristics to $\alpha$-induced scintillation because of the high
ionization density of the proton and triton tracks.

In establishing the picture described above, a study of scintillation
in the presence of an electric field played an important role. The
effect of an electric field on the LHe scintillation was first studied
by Hereford and Moss~\cite{HER66}. In this study, the authors measured
both the intensity of $\alpha$-particle-induced scintillation and the
ionization current extracted from the $\alpha$ tracks in the presence
of an electric field (9 to 43~kV/cm) in the temperature range of 1.23
to 4.2~K. The scintillation signal was integrated over $\sim
1$~$\mu$s. They observed a decrease in the scintillation intensity and
a corresponding increase in the ionization current with an increasing
electric field\footnote{Hereford and Moss~\cite{HER66} arrived at a
  conclusion that recombination luminescence accounts for
  approximately 60\% of the total scintillation intensity with zero
  electric field. While their experiment clearly established that
  recombination luminescence accounts for a large fraction of LHe
  scintillation, their analysis that led to the specific value of 60\%
  for the fraction due recombination luminescence does not appear
  valid in the light of what is now known about the properties of ions
  and excimers along the track of an $\alpha$ particle.}.

We have extended their work and have studied the effect of an
application of an electric field on both the prompt scintillation and
afterpulses in the temperature range of 0.2~K to 1.1~K with an
electric field up to $\sim 45$~kV/cm. We measured the time dependence
of the scintillation signal up to 14~$\mu$s.

The research reported here is directed primarily towards an
understanding of the physical processes underlying the electric field
dependence of the scintillation yield of liquid helium.  It provides
information on the dynamics of the charged particles generated by the
passage of an alpha particle through helium and the interaction of
molecular species produced on recombination. However, the work is
motivated by the desire to use helium scintillation in the presence of
a high electric field in the nEDM experiment~\cite{nEDM,ITO07}.

The nEDM experiment, currently being developed to be mounted at the
Spallation Neutron Source at the Oak Ridge National Laboratory, will
search for the neutron electric dipole moment (EDM) using a method
proposed by Golub and Lamoreaux~\cite{GOL94}. In this experiment, spin
polarized ultra cold neutrons will be produced and stored in a volume
filled with superfluid helium, to which spin polarized $^3$He atoms
will be added as cohabiting magnetometer. LHe scintillation produced
by the reaction products of the spin-dependent $^3$He($n$,$p$)$^3$H
reaction will be used to analyze the spin of the neutrons. The
signature of a non-zero EDM is a shift in the neutron precession
frequency upon application of an electric field. Therefore, the
effects of an electric field on LHe scintillation produced by heavy
particles such as protons, tritons, and $\alpha$ particles are of
particular interest. An earlier study performed by part of the nEDM
collaboration on LHe scintillation produced by the products of neutron
capture on $^3$He was reported in Ref.~\cite{ARC06}.

\section{Experimental apparatus and procedure}
\subsection{Overview}
The experiment was performed using a Leiden Cryogenics model CF-600
cryogen-free dilution refrigerator~\cite{LEIDEN}, which had a measured
cooling power of 1.8~mW at 0.2~K. A measurement cell, which housed
the electrodes immersed in LHe, was mounted on the mixing chamber of
the dilution refrigerator (DR). An $^{241}$Am $\alpha$ source was
electroplated on the ground electrode. $\alpha$ particles from the
source induced scintillation light in the volume of LHe between the
electrodes. The EUV scintillation light was converted to blue light by
a wavelength shifter, then was guided through a light guide to the
viewport at the bottom of the measurement cell and to a
PMT mounted outside the measurement cell in vacuum.
A description of each component is given below.

\subsection{Measurement cell, the electrodes, and the PMT}
\begin{figure}
\includegraphics[width=9cm]{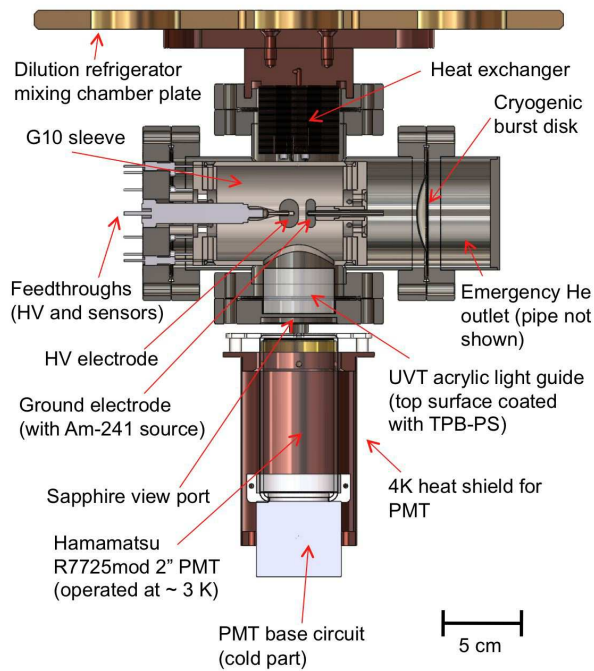}
\caption{(Color online) Schematic of the apparatus}
\label{fig:apparatus}
\end{figure}

A schematic of the measurement cell and the PMT is shown in
Fig.~\ref{fig:apparatus}. The measurement cell was made of a stainless
steel cross with 4-5/8'' and 4-3/4'' conflat flanges attached to its
ends. The volume of the LHe inside the cell was approximately
600~ml. At the top of the cell was a heat exchanger made of stack of
gold-coated copper plates, which provided cooling from the mixing
chamber to the LHe inside the cell.

The electrodes were made of stainless steel. The ground electrode was
19.05~mm in diameter and 6.35~mm in thickness, and had an edge rounded
with a radius of 3.175~mm. The high voltage (HV) electrode was
19.05~mm in diameter and 12.7~mm in thickness, and had an edge rounded
with a radius of 6.35~mm. The gap between the electrodes was
approximately 4~mm.

The HV electrode was mounted on a commercially made HV vacuum
feedthrough. The HV was supplied by a Glassman High Voltage Inc. Model
EH30R3 power supply. A thin wall stainless steel tubing was used as
the HV wire inside the cryostat to reduce the heat load. A HV resistor
($R=1$~G$\Omega$ or $R=1$~M$\Omega$) needed to be inserted in series
in the HV cable, in order to reduce the noise on the PMT signal
induced by the ripple of the HV power supply for the HV electrode.
The capacitance of the HV cable and this resistor formed a low pass
filter, which reduced the effect of the ripple to a sufficiently low
level.

Approximately 300~Bq of $^{241}$Am was electroplated on the ground
electrode. The activity had a diameter of 6~mm. Because the range of
5.5~MeV $\alpha$ particles in superfluid liquid helium is only $\sim
0.3$~mm, the ionization due to $\alpha$ particles occurred in a region
very close to the surface of the ground electrode. The electric field
at the location of the ionization was uniform to 5\%. Larger
electrodes and a smaller gap would have given a more uniform electric
field at the location of the ionization. However, that would have, at
the same time, significantly reduced the fraction of EUV photons
detected and hence the detection efficiency by limiting the solid
angle subtended at the location of the ionization by the
wavelength-shifter-coated surface. The size and shape of the
electrodes were chosen as a compromise between electric field
uniformity and the detection efficiency.

The electrodes were enclosed in a sleeve made of G10. This was to
prevent possible HV breakdowns due to the HV electrode seeing
structures on the inner walls of the stainless steel measurement
cell. 

A sapphire viewport with a view diameter of 49.3~mm was mounted at the
bottom of the measurement cell. An acrylic light guide with a diameter
of 49.3~mm and a length of 30.5~mm was mounted between the viewport
and the G10 sleeve. The G10 sleeve had a 48~mm diameter hole on the
side to allow the scintillation light to reach the light guide. The
top surface of the light guide was coated with polystyrene doped with
tetraphenyl butadiene (TPB), which is known to have good EUV-visible
conversion properties (see Ref.~\cite{MCK97} and references
therein). A Hamamatsu R7725 PMT, modified for cryogenic use with a Pt
underlayer~\cite{MEY10}, was mounted underneath the viewport in
vacuum.

A fraction of the $\alpha$ particle induced EUV light emitted from the
region between the electrodes was converted to blue light at the top
surface of the light guide. The blue light was then guided towards the
PMT through the light guide, transmitted through the sapphire viewport
and then detected by the PMT. The fraction of the EUV light that
reached the TPB coated surface of the light guide was about 5\%
(determined by the solid angle subtended by the coated surface of the
light guide at the location of scintillation events). The overall
detection efficiency of the apparatus (= the number of detected PEs
per emitted EUV photon) was estimated to be $\sim 3\times 10^{-4}$
from the solid angle mentioned above (5\%), the EUV to visible
conversion efficiency ($\sim 30$\%~\cite{MCK97}), the estimated
transport efficiency of the visible light ($\sim 10$\%), and the PMT's
quantum efficiency (18\%~\cite{MEY10}). The observed number of PEs in
the prompt pulse in the absence of an electric field $\sim 10.5$ (see
Sec.~\ref{sec:ADCanalysis}) was consistent with the expectation based
on the known scintillation efficiency (= the number of EUV photons in
the prompt pulse per energy deposition)~\cite{ADA95} and the estimated
detection efficiency.

The PMT was thermally anchored to the 3~K plate of the DR, which was
cooled by a pulse-tube refrigerator, in order to reduce heat load,
both radiative and conductive, to the measurement cell. The base
circuit adopted the split design described in Ref.~\cite{MEY10} with
some modifications. The resistor chain was heat-sunk on the 3~K
plate. In order to reduce the heat load from the Joule heating of the
resistor chain, a resistor value 10 times larger than the manufacturer
recommended value was used. Because of the low event rate ($\sim
300$~s$^{-1}$), this did not affect the performance of the PMT. The
rest of the base circuit, which consisted mostly of capacitors, was
mounted at the back of the PMT.  Previously, this PMT model had been
demonstrated to function properly at temperatures down to $\sim
8$~K~\cite{MEY10}. The present work showed that it functions properly
at $\sim 2.5$~K.

There were two RuO$_2$ temperature sensors mounted on the G10 sleeve
to monitor the temperature of the LHe inside the measurement
cell. There were two LHe level sensors inside the measurement
cell. They were each made of two concentric cylinders that formed a
capacitor. One was mounted in such way that it surrounded the heat
exchanger and the other surrounded the light guide, monitoring the LHe
level at the top and the bottom of the measurement cell. In addition
there was a LHe pressure sensor monitoring the LHe pressure inside the
measurement cell. A silicon diode sensor was mounted on the PMT as
well as on the PMT heat shield to monitor the temperature of the PMT
and the heat shield.

\subsection{Helium gas/liquid handling}
\label{sec:gashandling}
\begin{figure}
\includegraphics[width=9cm]{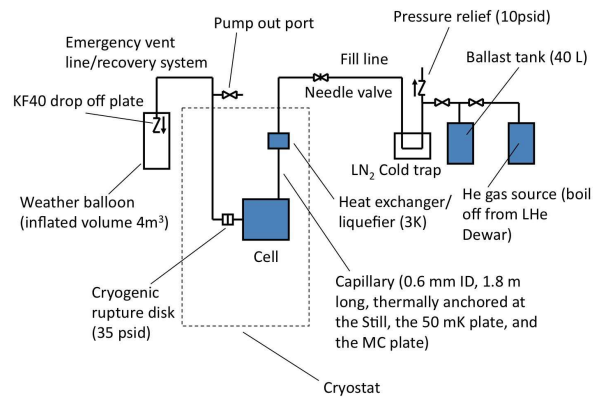}
\caption{(Color online) Flow schematic}
\label{fig:flowdiagram}
\end{figure}

A schematic of the helium gas/liquid flow in the system is shown in
Fig.~\ref{fig:flowdiagram}.  In filling the system, clean helium gas
taken from boiloff from a Dewar was further cleaned by a LN$_2$ cold
trap and then was introduced into the system. The helium was first
condensed by a liquefier (a box filled with sintered metal) mounted at
the 3~K plate, then flowed through a capillary tube with an ID of
0.6~mm and a length of 1.8~m which was thermally anchored at the
still, the 50~mK plate, and at the mixing chamber plate to the
measurement cell. The cell was filled with the DR running, which kept
the cell temperature at $\sim 0.9$~K during the fill. It took
approximately 24 hours to fill the 600 ml cell in this way.

\subsection{Data acquisition system}
\begin{figure}
\includegraphics[width=9cm]{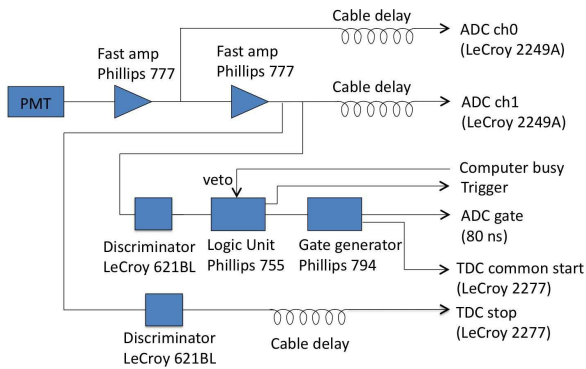}
\caption{(Color online) Schematic diagram of the DAQ system}
\label{fig:DAQdiagram}
\end{figure}

Data on the PMT signal were collected using a CAMAC and NIM based data
acquisition (DAQ) system. The schematic of the DAQ system is shown in
Fig.~\ref{fig:DAQdiagram}. The trigger was generated by a
discriminator whose threshold was set to a level corresponding to a
fraction of a PE, which allowed us to monitor the PMT gain using
single PE events from the dark noise.

\begin{figure}
\includegraphics[width=9cm]{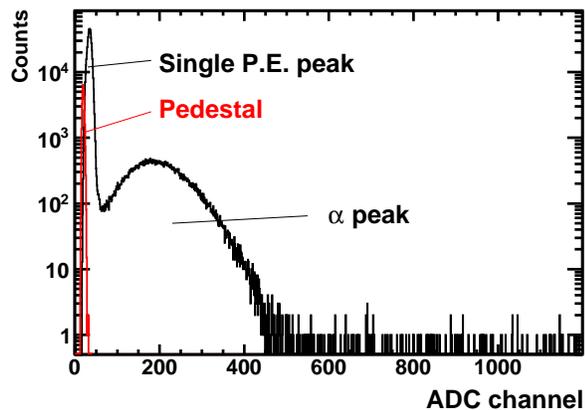}
\caption{(Color online) ADC spectrum of the prompt signal}
\label{fig:ADC}
\end{figure}

For each event, the size of the prompt signal was recorded using a
Lecroy 2249A charge-sensitive analog-to-digital converter (ADC) with a
80~ns-long gate. The length of the gate was chosen to capture all of
the prompt part of the signal (see Fig.~\ref{fig:PMTsignal}). The PMT
signal was amplified and then was split into two, one of which was
further amplified with a gain of $\sim 7$ before being fed to the
ADC. The high-gain signal ({\tt ADC ch1} in Fig.~\ref{fig:DAQdiagram})
was used to monitor the PMT gain using the single PE events from the
dark noise. The low-gain signal ({\tt ADC ch0} in
Fig.~\ref{fig:DAQdiagram}) was used to measure the size of the signal
generated by $\alpha$ particles. Randomly generated pulses were mixed
into the trigger in order to monitor the ADC pedestal. A typical
low-gain ADC spectrum is shown in Fig.~\ref{fig:ADC}.

\begin{figure}
\includegraphics[width=8cm]{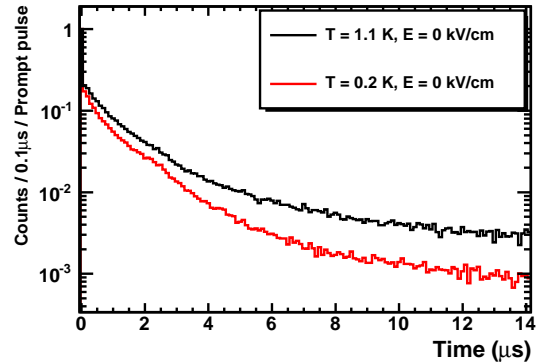}
\caption{(Color online) Spectra of afterpulse time of occurrence with respect to the
  prompt pulse. The bump around 2.5~$\mu$s is due to afterpulses
  caused by helium contamination in the PMT.}
\label{fig:TDC}
\end{figure}

\begin{figure}
\includegraphics[width=9cm]{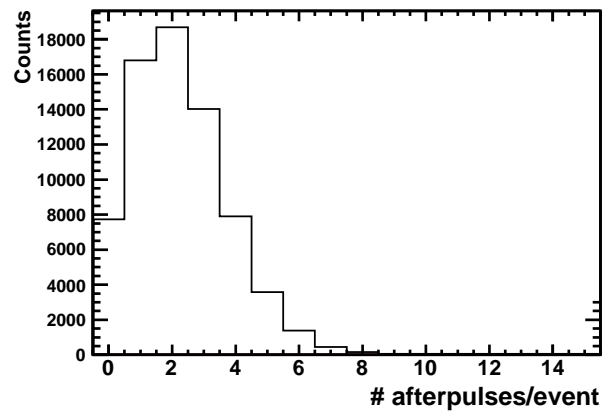}
\caption{Distribution of the number of afterpulses per event}
\label{fig:TDC_mult}
\end{figure}

The intensity of the delayed component was measured by recording the
timing of each afterpulse (see Fig.~\ref{fig:PMTsignal}) with respect
to the prompt pulse using a Lecroy 2277 multi-hit TDC. This TDC can
record up to 16 hits within a 16~$\mu$s time window. Typical time
spectra thus obtained are shown in Fig.~\ref{fig:TDC} for two
temperatures. Also a typical distribution of the number of afterpulses
per event is shown in Fig.~\ref{fig:TDC_mult}.

\subsection{Experiment}
The cell was filled with LHe following the procedure described in
Sec.~\ref{sec:gashandling}. Once the cell was filled and a desirable
measurement cell temperature was reached, the prompt scintillation
yield and the time spectrum of the afterpulses were measured as the
function of the applied HV. When increasing or decreasing the HV, the
PMT was kept turned off to protect the PMT from being exposed to
bright flashes of light due to HV breakdowns, which occurred
occasionally when changing the HV setting; we had found that exposing
the PMT to flashes of light from HV breakdowns decreased the PMT gain
and the change appeared to be permanent. The HV electrode was biased
with a negative voltage.  For each HV and temperature setting, data
were acquired for approximately 5 minutes, which corresponds to $\sim
10^5$ prompt scintillation events. Each of such a continuous data
taking period is called a ``run''.

Data were taken at five different temperature settings. The average
temperature and the 1-$\sigma$ variation in temperature for each of
these settings were $220 \pm 3$~mK, $298 \pm 3$~mK, $397 \pm 7$~mK,
$979 \pm 12$~mK, and $1084 \pm 17$~mK. In the remainder of this paper,
for conciseness we refer to these temperatures by their ``nominal''
values: 0.2~K, 0.3~K, 0.4~K, 1.0~K, and 1.1~K. In addition, data were
taken at temperatures up to 2~K with no HV applied. The gap in the
data between 0.4~K and 1.0~K was due to a thermal short that developed
in the system after the data were taken for 0.4~K, 0.2~K, and
0.3~K. The thermal short did not allow the system to run below 1.0~K.

As mentioned earlier, a HV resistor was inserted in the HV supply line
to suppress noise due to the ripple on the HV. Most of the data were
taken with a 1~G$\Omega$ resistor. The results did not change when a
1~M$\Omega$ resistor was used, confirming that the resistance between
the two electrodes was significantly larger than 1~G$\Omega$ and the
voltage across the electrode gap was indeed what was supplied by the
HV power supply.

\section{Data analysis and results}
\subsection{Analysis of the ADC data}
\label{sec:ADCanalysis}
The effect of the electric field on the prompt scintillation yield
manifests itself as a shift in the $\alpha$ peak in the low-gain ADC
spectrum. In order to remove effects due to possible changes in the
PMT gain, the horizontal scale of the spectrum needs to be properly
calibrated in terms of the number of PEs. As noted earlier, the gain
of the PMT was continuously monitored throughout the experiment by the
high-gain ADC that recorded single PE events from the dark noise. In
order to calibrate the horizontal scale of the low-gain ADC spectrum
using the PMT gain information obtained by the high-gain ADC, it is
necessary to determine the gain of the amplifier. This was achieved by
determining the slope of the correlation between the high-gain and
low-gain ADC spectra. Determination of the PMT gain and
the amplifier gain was performed for each run.

\begin{figure}
\includegraphics[width=9cm]{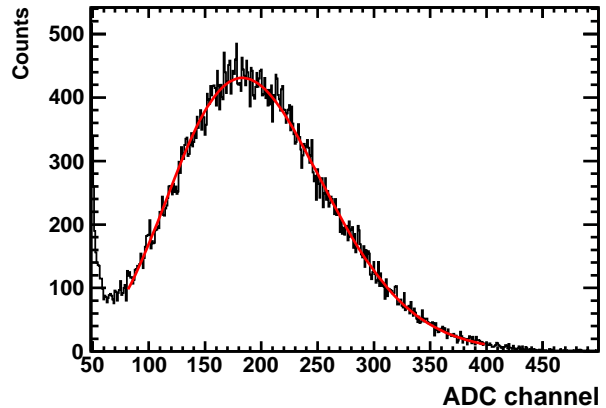}
\caption{(Color online) The $\alpha$ peak in the low-gain ADC spectrum fit with the model
  function in Eq.~(\ref{eq:ADCfit}).}
\label{fig:ADCfit}
\end{figure}

The resulting low-gain ADC spectrum was fit to determine the location of
the $\alpha$ peak using the following model function (a Poisson
distribution representing the distribution of the number of PEs
convoluted with a Gaussian distribution representing the PMT
response):
\begin{equation}
\label{eq:ADCfit}
f_{\rm ADC}(x)=N \sum_{k=1}^{\infty}\frac{\mu^ke^{-\mu}}{k!}\frac{1}
{\sqrt{2\pi{\sigma_k}^2}}e^{-\frac{(x-Gk)^2}{{\sigma_k}^2}}, 
\end{equation}
where $N$ is the overall normalization, $x$ a variable corresponding
to the ADC channel, $\mu$ the mean number of photoelectrons, and $G$
the gain of the system (the ADC channel corresponding to one
PE). $\sigma_k$ is given by
\begin{equation}
\sigma_k = \sqrt{{\sigma_{\rm PMT}}^2\cdot k+{\sigma_{\rm ped}}^2},
\end{equation}
where $\sigma_{\rm PMT}$ is the width of the response function of the
PMT for single PE events and $\sigma_{\rm ped}$ is the width of the
ADC pedestal. $\sigma_{\rm PMT}$ and $\sigma_{\rm ped}$ were
determined by fitting to the single PE and pedestal peaks respectively,
prior to fitting Eq.~(\ref{eq:ADCfit}) to the $\alpha$ peak. $N$ and
$\mu$ were varied in the $\alpha$ peak fit. The shape of the $\alpha$
peak in the low-gain ADC spectrum is well described by the model
function in Eq.~(\ref{eq:ADCfit}) (see Fig.~\ref{fig:ADCfit}).

\begin{figure}
\includegraphics[width=9cm]{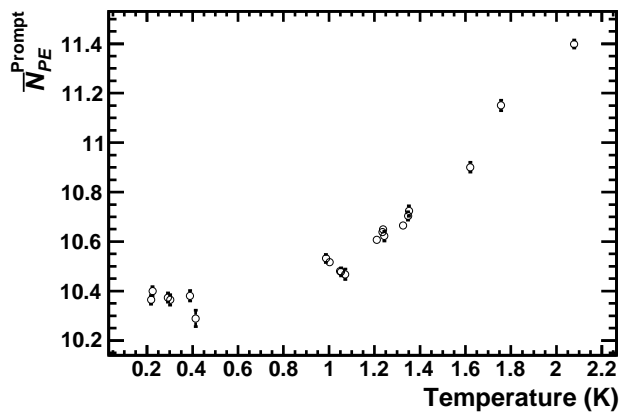}
\caption{The mean number of PEs observed in the prompt pulse
  ($\overline{N}_{PE}^{\rm prompt}$) vs the temperature.}
\label{fig:PEvstemp}
\end{figure}

\begin{figure}
\includegraphics[width=9cm]{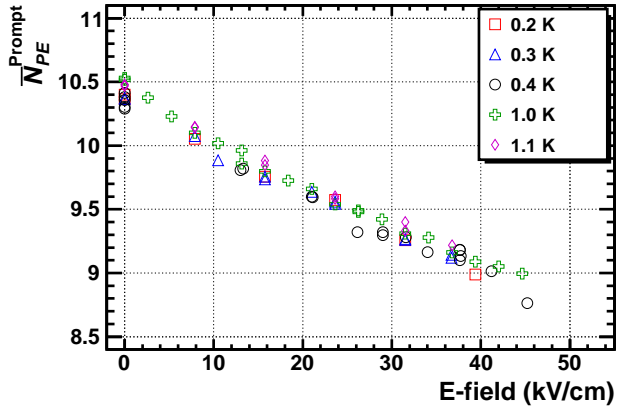}
\caption{(Color online) The mean number of PEs observed in the prompt pulse
  ($\overline{N}_{PE}^{\rm prompt}$) vs the electric field strength.}
\label{fig:PEvsHV}
\end{figure}

%
Figure~\ref{fig:PEvstemp} shows the dependence of the mean number of
PEs observed in the prompt pulse ($\overline{N}_{PE}^{\rm prompt}$) on
the temperature with no electric field. Shown in Fig.~\ref{fig:PEvsHV}
is $\overline{N}_{PE}^{\rm prompt}$ plotted against the strength of
the electric field for different temperatures. The prompt
scintillation yield decreases by about 15\% with an electric field of
45~kV/cm. The effect of the electric field on the prompt scintillation
has little temperature dependence.

Our data shown in Figs.~\ref{fig:PEvstemp} and \ref{fig:PEvsHV} are
qualitatively consistent with what was observed in the earlier work by
Hereford and coworkers~\cite{HER66,ROB73}. (Compare, for example,
Fig.~\ref{fig:PEvstemp} of this work with Fig.~2 of Ref.~\cite{ROB73}
and Fig.~\ref{fig:PEvsHV} of this work with Fig.~3 of
Ref.~\cite{HER66}.)  Note, however, that a direct comparison between
our data and the data in Refs.~\cite{HER66} and \cite{ROB73} cannot be
made because of the difference in the integration time for the prompt
pulse.

\subsection{Analysis of the TDC data}
\label{sec:AfterpulseAnalysis}
\begin{figure}
\includegraphics[width=9cm]{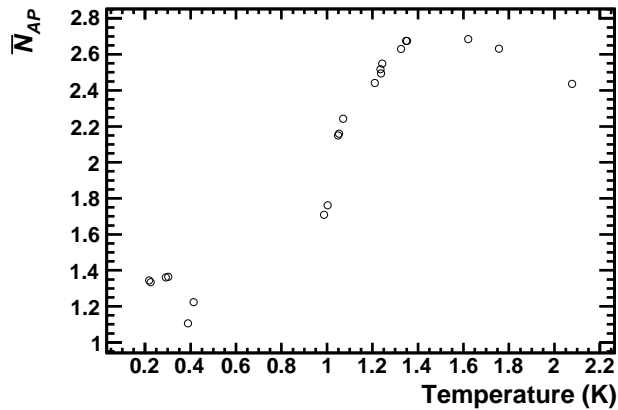}
\caption{The mean number of afterpulses per prompt pulse observed in
  the first 14~$\mu$s ($\overline{N}_{AP}$) plotted as a function of
  the temperature for zero electric field.}
\label{fig:APvsT}
\end{figure}

The mean number of afterpulses per prompt pulse observed in the first
14~$\mu$s was determined by fitting a Poisson distribution to the
observed distribution of the number of afterpulses per event
(Fig.~\ref{fig:TDC_mult}). The mean number of afterpulses per prompt
pulse thus obtained ($\overline{N}_{AP}$) is plotted against the
temperature in Fig.~\ref{fig:APvsT}. Our results, showing a reduction
in the afterpulse intensity at lowered temperatures, are consistent
with Refs.~\cite{MCK03} and \cite{ROB73}. (Although Ref.~\cite{ROB73}
only shows a plot of the ``pulse height'' which is the signal
integrated for the first 1~$\mu$s, and the total intensity plotted
against the temperature, the temperature dependence of the afterpulse
intensity can be inferred from the data.)

\begin{figure}
\includegraphics[width=9cm]{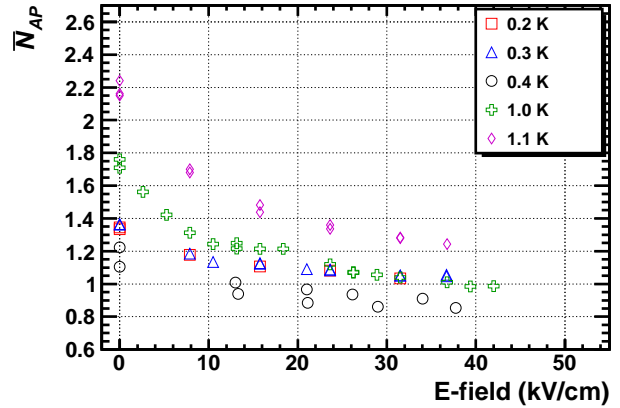}
\caption{(Color online) The mean number of afterpulses per prompt pulse observed in
  the first 14~$\mu$s ($\overline{N}_{AP}$) plotted against the
  strength of the electric field.}
\label{fig:APvsE}
\end{figure}

\begin{figure}
\includegraphics[width=9cm]{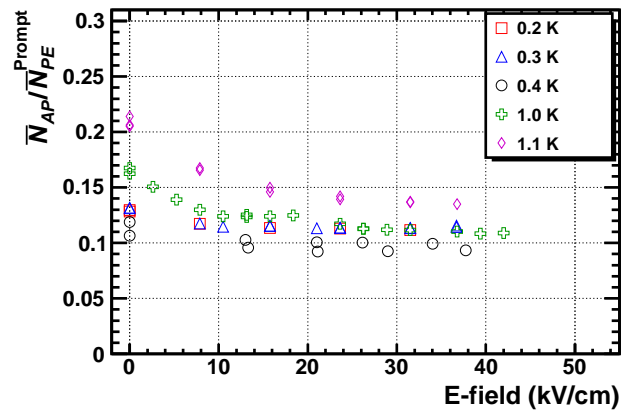}
\caption{(Color online) The mean number of the afterpulses normalized to the mean
  number of PEs observed in the prompt pulse
  ($\overline{N}_{AP}/\overline{N}_{PE}^{\rm prompt}$) plotted against
  the strength of the electric field.}
\label{fig:APperPE}
\end{figure}

Figures~\ref{fig:APvsE} and \ref{fig:APperPE} show the mean number of
the afterpulses ($\overline{N}_{AP}$) and the mean number of the
afterpulses normalized to the mean number of PEs observed in the
prompt pulse ($\overline{N}_{AP}/\overline{N}_{PE}^{\rm prompt}$)
plotted against the strength of the electric field, respectively. We
see from the figures that the afterpulse intensity decreases with an
increasing electric field strength (Fig.~\ref{fig:APvsE}) and that the
afterpulse intensity is more strongly affected by the electric field
than the prompt scintillation (Fig.~\ref{fig:APperPE}). Also, in
contrast to the prompt scintillation, the effect of the electric field
on the afterpulse intensity exhibits some temperature dependence.


\section{Discussion}
What we have observed from our data can be summarized as follows:\\
1) The prompt scintillation yield is reduced by 15\% at $\sim 45$~kV/cm.\\
2) The reduction in the prompt scintillation yield is approximately
  linear in the strength of the electric field.\\
3) The effect of the electric field on the prompt scintillation yield 
  has a very weak temperature dependence in the temperature range of
  0.2~K to 1.1~K.\\
4) The electric field has a stronger effect on the intensity of the
  delayed component than it has on the intensity of the prompt pulse.\\
5) The observed temperature dependence of the prompt and
    afterpulse intensities at zero electric field is in qualitative
    agreement with previous work.\\

Below we discuss the implication of these observations.

\subsection{Effect of an electric field on the prompt scintillation}
\label{sec:promptdiscussion}
The prompt scintillation signal is the result of the radiative decay
of singlet He$^*$ atoms in excited states and of He$_2^*$ excimers
also in singlet states. An $\alpha$ particle produces these excimers
and atoms either by ionization followed by recombination or by
promoting an atom directly to an excited state without ionization. The
ratio of the number of direct excitations to ionizations has been
predicted in helium to be 0.45:1~\cite{SAT74}. (Although this
calculation was performed with 100~keV electrons as the primary
particle, the expectation is that it holds for $\alpha$ particles as
well~\cite{SAT76}.)  Among the atoms promoted directly to excited
states, the ratio of singlets to triplets was calculated to be 5:1. On
the other hand, the ratio of singlets to triplets formed by
recombination of ionized atoms is expected to be the statistical value
of 1:3 for $\alpha$ particles (not the case when electrons are the
primary ionizing particle~\cite{ADA01}).  Scintillation from atoms in
excited states that have previously not undergone ionization is not
affected by an electric field whereas the number of those that have
been ionized is decreased in an electric field because of charge
separation. Were there no other complicating factors, the electric
field-dependent fraction, $x$, of the prompt scintillation would be $x
=0.4$. However, a number of phenomena can influence this fraction,
most likely increasing its value.

1) The prompt signal for $\alpha$ particles is substantially
quenched~\cite{ADA01} by the nonradiative destruction of singlet
excimers by the Penning process [Eq.~(\ref{eq:penning2})] but by the
same mechanism it may be slightly enhanced with the destruction of
triplet excimers and the subsequent recombination of the resulting
ions into singlet states.  Excited state singlet atoms may also
nonradiatively decay upon interaction with other species of
excitations. (Nonradiative destruction of singlet species is discussed
in Sec.~\ref{sec:quenching}.)

2) Singlet helium atoms in excited states with principal quantum 
number, $n$, of 3 or greater can autoionize by the Hornbeck-Molnar 
process~\cite{HOR51}
\begin{equation}
\label{eq:hornbeck}
{\rm He}^* + {\rm He} \rightarrow {\rm He}_2^+  + e^-,
\end{equation}
since the binding energy of He$_2^+$ is $\sim 2$~eV, which is greater
than the energy to ionize a He($n\geq 3$) atom. Based on the
oscillator strengths for the transitions between the ground state and
the various excited states of helium~\cite{BER97} slightly more than
one third of the atoms promoted to excited states will have a
principal quantum number of 3 or greater, the other two thirds having
$n = 2$.  If all atoms with $n\geq 3$ were to undergo autoionization,
then the electric field-dependent fraction of the prompt scintillation
would be increased to $x =0.6$.


Various theoretical models have been developed that describe
electron-ion recombination in an ionization track in fluid media:
Jaffe's columnar theory of recombination~\cite{JAF13,KRA52} describes
the case in which a dense plasma of positive and negative ions is
formed along the ionization tracks while Onsager's theory~\cite{ONS38}
describes the case where each electron-ion pair is spatially separated
and recombination occurs between an electron and its parent ion
(geminate recombination).  For the highly ionizing track of an
$\alpha$ particle in LHe, the Jaffe model would appear the more
applicable.

In the columnar theory, the rate of recombination is governed by the
following equations:
\begin{equation}
\label{eq:nplus}
\frac{\partial n_+}{\partial t} = -u_+ {\pmb E}\cdot {\pmb \nabla}n_+ +
D_+ \nabla^2n_+ - \alpha n_-n_+,
\end{equation}
and
\begin{equation}
\label{eq:nminus}
\frac{\partial n_-}{\partial t} = u_- {\pmb E}\cdot {\pmb \nabla}n_- +
D_- \nabla^2n_- - \alpha n_+n_-,
\end{equation}                                          
where $n_+$ and $n_-$ are the densities of the snowballs and bubbles,
$u_+$ and $u_-$ are the mobilities, $D_+$ and $D_-$ are the diffusion
coefficients, and $\alpha$ is the recombination coefficient. What this
model describes are clouds of negative and position ions being pulled
away from each other by the electric field while at the same time
recombining at a rate proportional to the product of their
densities. Also, the clouds expand radially due to the diffusive
motion of the charges. Both the reduction in the scintillation yield
and the generation of an ionization current result from the suppression
of recombination due to the presence of an electric field. The motion
and recombination of electrons and ions described by these equations
occur on a time scale of $\sim 10^{-10}$~s after an $\alpha$ particle
is stopped in helium.  The processes associated with the decay and
destruction of excimers created on recombination, processes that are
not influenced by the electric field and occur on time scales of
$10^{-9}$ to $10^{-8}$~s, are discussed separately in later sections.

\begin{figure}
\includegraphics[width=9cm]{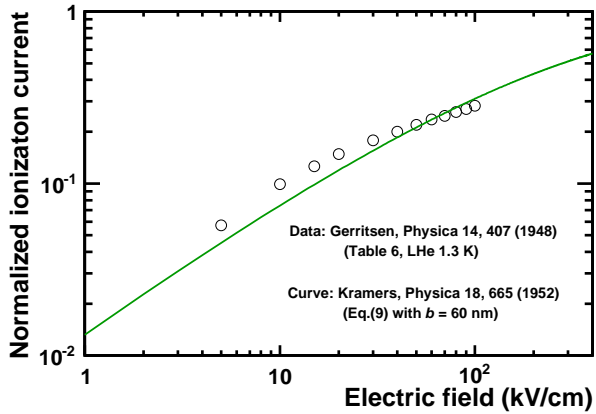}
\caption{(Color online) The best fit of Kramers' theory to Gerritsen's data with $b =
  60$~nm. The ionization current is normalized to the saturation
  current.}
\label{fig:gerritsen}
\end{figure}

In looking for a solution to these rate equations Jaffe assumed the
diffusion terms were larger than those associated with recombination
and hence, along with the electric field term, were the principal
cause of the change in densities of the charges.  This is a valid
approach for ionization in gases. On the other hand,
Kramers~\cite{KRA52}, in attempting to explain the electric field
dependence of the ionization currents from $\alpha$ particles stopped
in liquid helium measured by Gerritsen~\cite{GER48}, pointed out that
in a dense fluid the diffusivity of the ions is small and that
recombination has a much larger influence on the time dependence of
the charge density. Gerritsen's data for the electric field dependence
of the ionization current can be fit approximately by the Kramers
theory with a cylindrical Gaussian charge distribution having the same
density and spatial distribution for $n_+$ and $n_-$
\begin{equation}
n_{\pm}(t=0) = \frac{N_0}{\pi b^2 }e^{-r^2/b^2},
\end{equation}
where $N_0$ is the initial number of positive ion snowballs or
electron bubbles per unit length along the track. The diffusion
coefficient was assumed to be zero. The relationship between the
recombination coefficient and the mobilities is given by the Langevin
relation (see below). The best fit of Kramers' theory to Gerritsen's
data with the one adjustable parameter, $b = 60$~nm, is shown in
Fig.~\ref{fig:gerritsen}.

Several comments regarding this result are relevant to understanding the 
scintillation and ionization currents of LHe.

1) The ionization current data contains contributions that are not
included in the Jaffe-Kramers theory, which describes the ionization
current due to charge carriers that escaped the initial
recombination~\footnote{In the literature, the term ``initial
  recombination'' is used to mean ``geminate recombination'' as
  discussed in Ref.~\cite{ONS38}. Here we use the term ``initial
  recombination'' to mean the recombination that immediately follows
  the passage of an $\alpha$ particle. }. However, at electric field
strengths below that corresponding to the saturation current, charge
carriers that initially recombine can contribute to the ionization
current through processes such as Penning ionization
[Eqs.~(\ref{eq:penning1}) and (\ref{eq:penning2})]. A crude estimate
indicates that such contributions could be larger than 10\%.

2) The Langevin relation between recombination coefficient and
mobility is $\alpha = e (u_+ + u_-)/ \epsilon_0$. In the approximation
that the mobilities of the two species are the same, the electric
field dependence of the current is independent of mobility, as it
enters the recombination and electric field terms in the same
manner. At the temperature within the track of approximately 2~K (see
below), the mobilities of the positive and negative ions are
approximately the same~\cite{DON98}, having the value of $u_+ + u_-
\sim 10^{-5}$~m$^2$/(V$\cdot$s). Hence the recombination coefficient
is $ \alpha \approx 2 \times 10^{-13}$~m$^3$/s.

3) From the Einstein relation $D=kTu/e$, the diffusion coefficient is
calculated to be $\sim 1.7 \times 10^{-9}$~m$^2$/s. Since the initial
charge densities along the track are the order of $10^{22}$~m$^{-3}$,
diffusion plays little role in affecting the density distribution at
early times when recombination is dominant. At longer times when the
concentrations have decreased substantially, diffusion can influence
the charge separation.

4) The assumption that the positive ions and the electron bubbles
initially have the same spatial distribution is unlikely to be
correct. The ionization events from the $\alpha$ particle or from
secondary electrons with energy sufficient to create further
ionizations occur in a small cylinder about the track of less than
10~nm. He ions do not move appreciably in the time it takes for the
electrons to thermalize and form a bubble. However, once the energy of
secondary electrons drops below 20~eV, the first excited state of He,
the only process by which they can lose energy is elastic scattering
from helium atoms. Since the fraction of energy an electron loses in
such a collision is very small, being dependent on the ratio of the
mass of the electron to that of the atom, the order of $10^4$
collisions are required to decrease an initial energy of 10~eV to
below 0.1~eV thought to be necessary for bubble formation. From the
cross section for elastic scattering, which varies from $3\ {\rm
  to}\ 6\times 10^{-16}$~cm$^2$ between 1 and 10~eV, and the liquid
density, the mean free path is roughly $\ell \sim 1$~nm and the mean
distance for diffusion, proportional to the square root of the number
of scatters, is the order of 100~nm~\cite{BEN99}.  But as Mu$\tilde{\rm
  n}$oz {\it et al.}~\cite{MUN86} point out, if the initial positive
and negative charge distributions are Gaussian with very different
radii, then the radial field arising from the different charge
distributions can be large, easily in the range of 100 to 1000~kV/cm
for an $\alpha$ track in helium assuming 10~nm for the radius of the
positive ions and 100~nm for the electrons. The two distinct
distributions will therefore rapidly merge to form a more uniform
distribution assumed in the Jaffe and Kramers theories. The positive
ion snowballs having a mobility comparable to that of the electron
bubble at the initial temperature of $\sim 2$~K along the track (see
below), will expand outward and the electrons contract inward.  

5) A significant fraction of the secondary electron produced by an
$\alpha$ particle have energies above the He ionization energy of
24.6~eV and are therefore capable of creating further ionization.  At
low energies the probability that a secondary electron will have a
particular energy $E$ decreases relatively slowly with increasing
energy, varying as $E^{-2}$~\cite{ROS52}. Since the cross section for
ionization or excitation by an electron in the range of 25 to several
hundred eV is the order of $10^{-17}$~cm$^2$, these
ionizations/excitations would occur a considerable distance from the
track were it not for elastic scattering. The elastic scattering cross
section for an 100~eV electron is $1\times 10^{-16}$~cm$^2$ so that
energetic electrons will undergo many random scatters, on average
remaining in the vicinity of the $\alpha$ track. Such electron may
have the effect of modifying the charge distribution but not to the
extent that a Gaussian distribution is not a reasonable approximation.

\begin{figure}
\includegraphics[width=9cm]{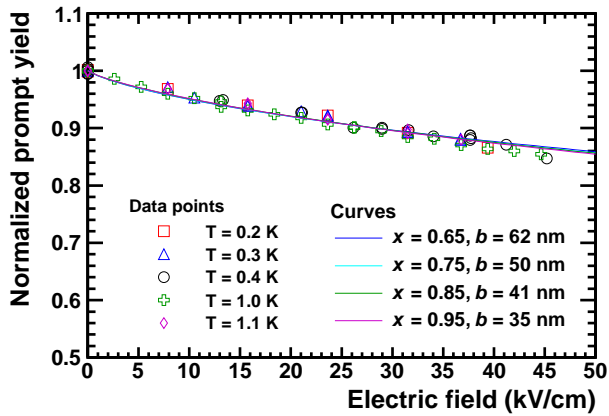}
\caption{(Color online) Kramers theory fit to the electric field strength dependence
  of the prompt scintillation yield measured by the current work. The
  prompt scintillation yield is normalized to the value at $E=0$. The
  curves are calculated using Eq.(9) of Ref.~\cite{KRA52}.}
\label{fig:PromptFit}
\end{figure}

\begin{figure}
\includegraphics[width=8cm]{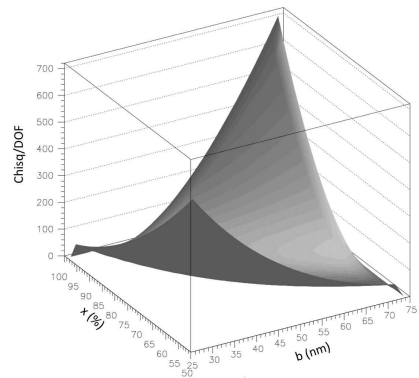}
\caption{$\chi^2$ per degree of freedom of the fit of Kramers theory
  to the prompt scintillation data plotted against $x$ and $b$.}
\label{fig:b_vs_x}
\end{figure}

The measured electric field dependence of the prompt scintillation can
be fit well using the Kramers theory for a range of different Gaussian
widths, $b$. The value of $b$ depends on the choice of the fraction,
$x$, of scintillation resulting from species created by ionization,
which can be affected by the electric field. The results are shown in
Fig.~\ref{fig:PromptFit}.  Figure~\ref{fig:b_vs_x} shows the $\chi^2$
per degree of freedom of the fit of Kramers theory to the prompt
scintillation data plotted against $x$ and $b$. If $x$ is taken to be
0.95 then the best fit is for $b =35$~nm, while for $x = 0.65$,
$b=62$~nm. In the Kramers theory~\cite{KRA52}, the fraction of ions
that recombine depends on a single parameter
$f=\sqrt{\pi}\epsilon_0 bE/(N_0e)$. That is, increasing the
radius of the track $b$ has the same effect as increasing the electric
field $E$. From this it is expected that a smaller $x$ would require
a larger $b$. Our results (Figs.~\ref{fig:PromptFit} and
\ref{fig:b_vs_x}) are consistent with this expectation.


A few comments about the relationship between ionization current and
the effect of an electric field on the prompt scintillation intensity
are appropriate. These two types of data reflect two different aspects
of the same phenomenon, and as such it should be possible to construct
a model that describe both data consistently. However, such an attempt
is complicated by a number of factors. 1) As mentioned earlier, the
ionization current has a contribution from charge carriers that
initially recombine. 2) There is uncertainty in the value of $x$.  3)
Even with a short integration time (in our case $\sim 80$~ns), there
is inevitably a contribution to the prompt signal from the delayed
component, which has a different electric field dependence (see Sec
IV.C). Because of these complications we have not attempted a unified
analysis of our data and Gerritsen's ionization current data. While we
note that, as seen in Figs.~\ref{fig:PromptFit} and \ref{fig:b_vs_x},
there is considerable range in the values of $x$ and $b$ that fit the
data, the choice of $x \approx 0.6$ and $b \approx 60$~nm is
consistent both with our estimate of the fraction $x$ and the best fit
of Kramers' theory to Gerritsen's data.

As seen in Fig.~\ref{fig:PromptFit}, the effect of the electric field
on the prompt scintillation has little temperature dependence. This is
attributed to three causes: 1) The effect of the electric field on
recombination is nearly independent of the mobility of the ions as
discussed earlier. 2) The initial temperature of the track is in the
vicinity of 2 K, irrespective of the temperature of the bulk liquid
(see below), resulting in the same mobility of ions in the track
irrespective of the bulk liquid temperature. 3) The characteristic
time of recombination $t_r = 1/(\alpha n_0)$, where $n_0$ is the
initial ionization density, is shorter than the time it takes the
$\alpha$ track to thermalize with the bulk liquid $t_{\rm therm} =
b/c$, where $c$ is the sound velocity ($c\sim 230$~m/s). Indeed, with
$b=60$~nm, which gives $n_0\sim 4\times 10^{22}$~m$^{-3}$, the
recombination time is $t_r\sim 1\times 10^{-10}$~s as compared to the
thermalization time $t_{\rm therm}\sim 3\times 10^{-10}$~s.

\subsection{Temperature dependence of the prompt scintillation}

The fraction of energy of an $\alpha$ particle that is quickly down
converted to excitations of the superfluid, phonons and rotons, is
sufficient to heat a column of liquid a few tens of nanometers radius
along the track to the vicinity of 2~K irrespective of how low the
temperature of the bulk liquid~\cite{BAN95}. Since the density of
liquid helium changes by less that 0.5\% below 2~K, the initial
environment in which the electrons, ions, excimers and excited state
atoms reside, on their creation, is independent of the temperature of
the bulk liquid. Therefore, the temperature dependence of the prompt
scintillation, illustrated in Fig.~\ref{fig:PEvstemp}, must result
from the manner in which those entities responsible for the
scintillation are influenced by the expansion of the roton/phonon
cloud (with velocity of $\sim 10^2$~m/s) taking place at times the
order of $10^{-10}$~s. Below a temperature of 0.6~K the density of
thermal excitations in the bulk liquid is so low that the hot columnar
cloud expands into essentially a vacuum. But at higher temperatures
the thermal excitations in the bulk become sufficiently dense to
scatter and slow the radially expanding cloud.

As discussed below, the $1/t$ component of the afterpulse intensity also 
decreases with decreasing temperature. The temperature dependence of both 
the prompt and afterpulse signals is attributed to the more rapid decrease of
the triplet excimer density due to expansion, leading to a reduced production 
of singlets via the Penning process.

Hereford and coworkers~\cite{ROB73,ROB71}, in modeling the temperature 
dependence of the scintillation they observed, argued that the 
increased diffusion at low temperatures results increased quenching, 
that is, the enhanced nonradiative destruction of the singlet species. 
However, their argument was not informed by the fact that the 
roton and phonon densities in the vicinity of the track is 
essentially independent of the temperature of the bulk liquid.  

\subsection{Temperature and electric field dependence 
of the afterpulse scintillation} 
Following the work of McKinsey {\it et al.}~\cite{MCK03} the time
dependence of the afterpulse scintillation for times between
0.4~$\mu$s and 14~$\mu$s has been characterized using the function
\begin{equation}
\label{eq:APspectrum}
f_{\rm TDC}(t)=Ae^{-t/\tau_s}+\frac{B}{t} + C.
\end{equation}
$\tau_s$, $A$, $B$, and $C$ were varied in the fit. For convenience,
we define $N_{AP}^{\rm Exp}$, $N_{AP}^{1/t}$, and $N_{AP}^{\rm Const}$ to be the contributions to the
number of afterpulses from the first, second, and third terms in
Eq.~\ref{eq:APspectrum}, respectively, that is
\begin{equation}
N_{AP}^{\rm Exp} = \int Ae^{-t/\tau_s} dt,
\end{equation}
\begin{equation}
N_{AP}^{1/t} = \int \frac{B}{t} dt,
\end{equation}
and 
\begin{equation}
N_{AP}^{\rm Const} = C \int dt.
\end{equation}

\begin{figure}
\includegraphics[width=9cm]{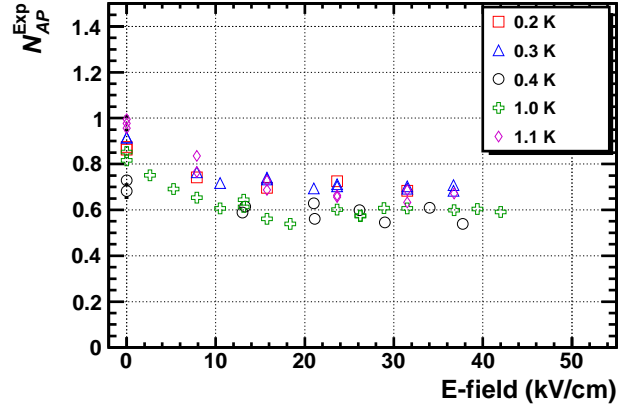}
\caption{(Color online) $N_{AP}^{\rm Exp}$, the number of afterpulses from the exponential
  component ($e^{-t/\tau_s}$)  in the afterpulse time
  spectrum plotted against the strength of the electric field.}
\label{fig:APexp}
\end{figure}

\begin{figure}
\includegraphics[width=9cm]{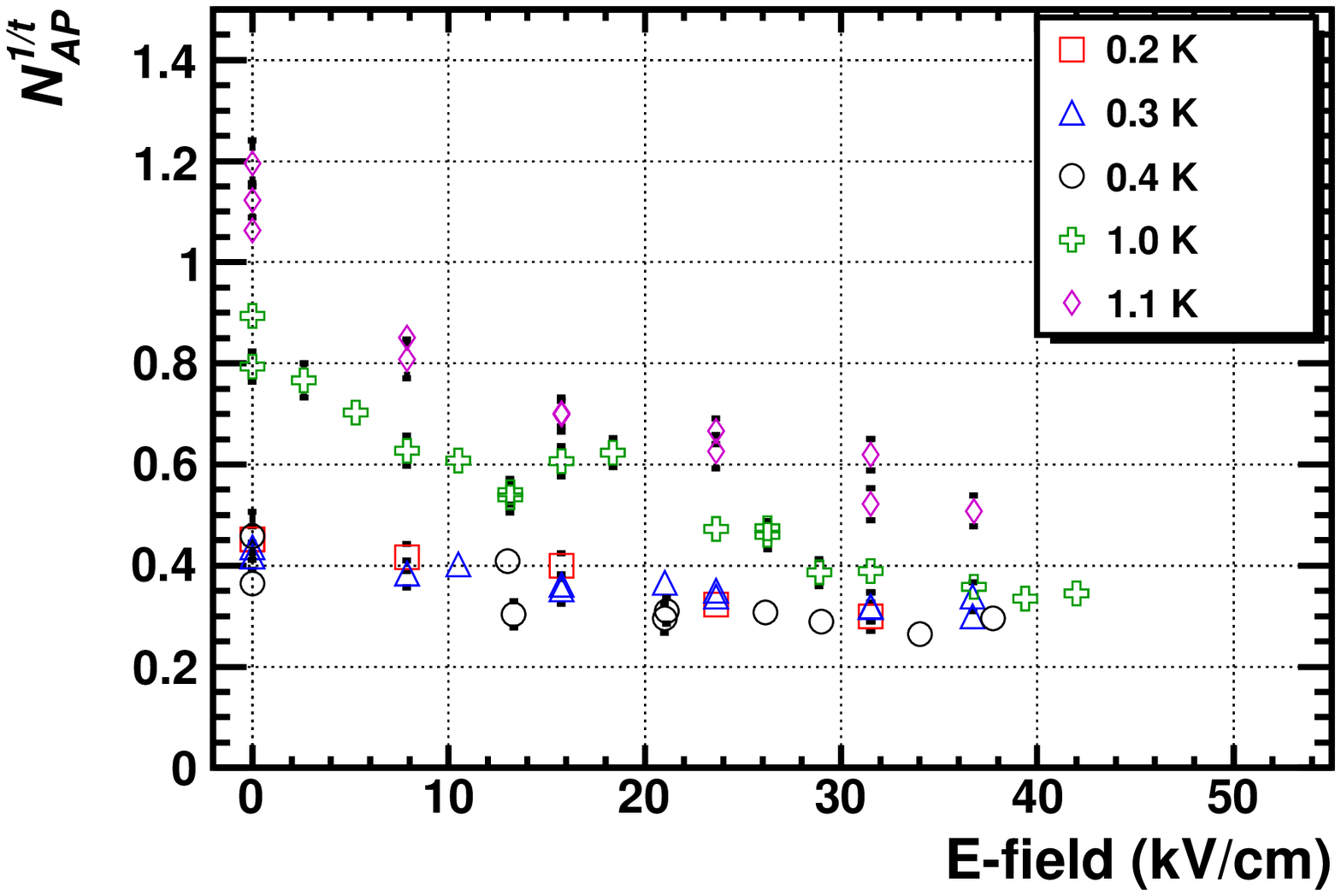}
\caption{(Color online) $N_{AP}^{1/t}$, the number of afterpulses from the $t^{-1}$ component in the afterpulse
  time spectrum plotted against the strength of the electric field.}
\label{fig:APtinv}
\end{figure}

\begin{figure}
\includegraphics[width=9cm]{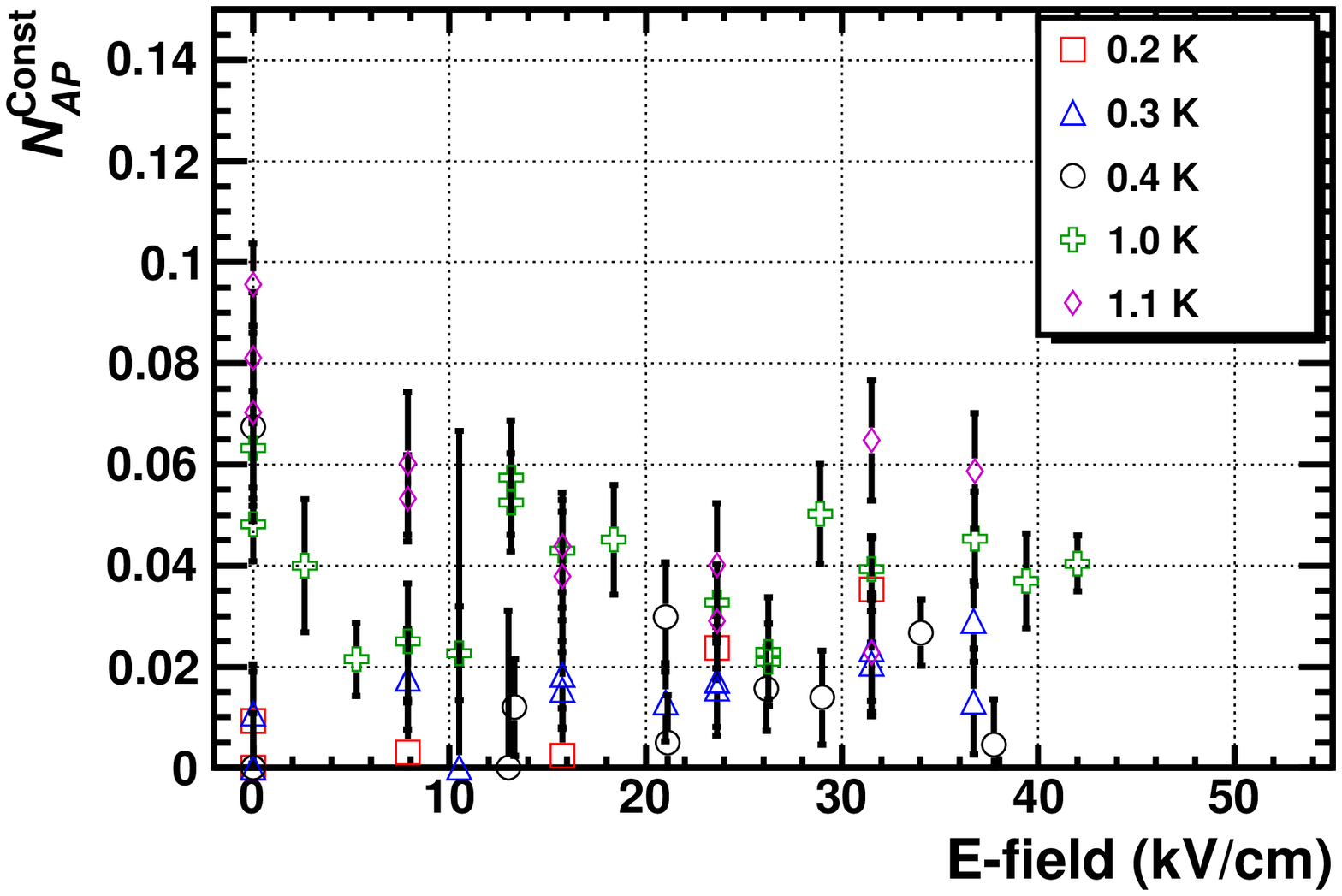}
\caption{(Color online) $N_{AP}^{\rm Const}$, the number of afterpulses from the constant component in the afterpulse
  time spectrum plotted against the strength of the electric field.}
\label{fig:APconst}
\end{figure}

$N_{AP}^{\rm Exp}$, $N_{AP}^{1/t}$, and $N_{AP}^{\rm Const}$ are
plotted as a function of the electric field for different temperatures
in Figs.~\ref{fig:APexp}, \ref{fig:APtinv}, and \ref{fig:APconst}
respectively. In Figs.~\ref{fig:APexp} and \ref{fig:APtinv}, the 1.0~K
data points for electric fields between 10~kV/cm and 20~kV/cm seem to
show a larger scatter. We are not able to offer an explanation.

The value of $\tau_s$ that we obtained was $\sim 1.3$~$\mu$s, slightly
shorter than the value quoted in Ref.~\cite{MCK03}. Our results show
that it is nearly independent of the temperature and the electric
field. 

\begin{figure}
\includegraphics[width=9cm]{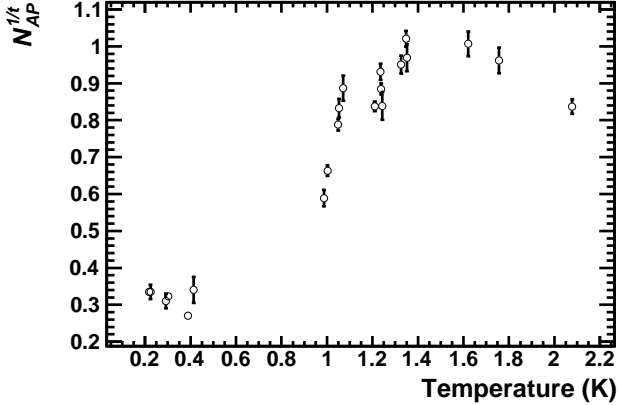}
\caption{$N_{AP}^{1/t}$, the number of afterpulses from the $t^{-1}$
  component in the afterpulse time spectrum plotted as a function of
  the temperature for zero electric field.}
\label{fig:AtinvvsT}
\end{figure}

Also in Fig.~\ref{fig:AtinvvsT}, $N_{AP}^{1/t}$ is plotted as a
function of the temperature for zero electric field.

At times longer than $10^{-7}$~s all singlet state excimers and
excited atoms (except, perhaps, for the first excited state
He(2$^1$S), which in vacuum has a lifetime of 19~ms~\cite{VAN71}),
will have long decayed. Any photons produced at a later time must
result from triplet states.

\subsubsection{Component dependent exponentially on time}  
McKinsey {\it et al.}~\cite{MCK03} hypothesized that the afterpulse
term dependent exponentially on time was the result of He(2$^1$S)
atoms forming excimers in A$^1\Sigma_u^+$ states from which they
radiatively decay to the dissociated ground state of the helium
pair. Given that the lifetime of the triplet state He(2$^3$S) has been
determined in the liquid to be 15~$\mu$s~\cite{KET74a}, this appears
to be a reasonable hypothesis.  Furthermore, there appears to be
little or no temperature dependence to the exponentially
time-dependent term, as expected for such a source.  On the other
hand, one would not expect that a signal arising from a neutral atom
to exhibit a dependence on electric field as we have observed (see
Fig. \ref{fig:APexp}), unless it originated through recombination.  Since
the probability that ions form excimers prior to recombination is
high, atomic He(2$^1$S) appears more likely to be produced by decay
from a higher lying He(n$^1$P) state created de novo by
electromagnetic excitation from the ground state than through a
cascade from an atom formed by recombination. We are unable to offer
an explanation of this exponentially time-dependent component of the
afterpulse scintillation.

\subsubsection{Component dependent inversely on time}
Keto {\it et al.}~\cite{KET74b}, using a beam of
160~keV electrons as the primary ionizing source, measured the 
long-time afterglow in the infrared spectra associated with a number 
of transitions between excited states 
 of excimers and atoms to have a time dependence that varied as
$1/t$. They ascribed this time dependence to the production of the
excited states by the Penning ionization of triplet excimers, 
since the bimolecular process,
\begin{equation}
\label{eq:birate}
\frac{dn}{dt} = - \gamma n^2
\end{equation}
results in $n$ proportional to $1/t$ for $\gamma t > 1$. A more
detailed description of the time dependence of the singlet excimer
density is expressed by Eq.~(\ref{eq:quench}) in the next Section. A
similar relation can be given for the triplet density, the essentials
of which for the purposes of this discussion are contained in
Eq.~(\ref{eq:birate}).

The problem of describing the time dependence of the afterpulse 
scintillation from an $\alpha$ particle differs from that in the work
of Keto {\it et al.}~\cite{KET74b} since in one case the density 
of excimers is uniform whereas in the other  the 
 excimer cloud is expanding about the $\alpha$ track.

King and Voltz~\cite{KIN66} developed a theoretical model of afterpulses
that includes the spatial evolution of the excimer density with time.
 The description involves the kinetics
of the diffusion of the triplets at times well after the prompt 
scintillation has decayed. If the triplets maintain a Gaussian 
distribution as they expand by diffusion, then at long times the 
afterpulsing decreases approximately inversely with time. McKinsey 
{\it et al.}~\cite{MCK03} cited this model by way of explaining the 
origin of the $1/t$ component of the afterpulses. At shorter times 
the King model predicts a time variation that is considerably
faster than $1/t$, but it does not provide an explanation of the
exponential dependence on time, as experimentally observed.

The amplitude of the $1/t$ afterpulse component exhibits interesting
temperature and electric field dependences. Firstly, the number of
afterpulses varies much more with temperature than the number of
photons in the prompt scintillation. Secondly, the number of
afterpulses decreases more rapidly with increasing electric field than
the prompt signal.

As with the slight variation of the prompt signal with temperature
below 1~K, the rapid variation of the $1/t$ afterpulse component with
temperature can be understood qualitatively as being the result of the
change in propagation of the triplet excimers away from the $\alpha$
track.  While the temperature within the core of the track is not
sensitive to the ambient temperature of the liquid, the thermal
excitations in the surrounding liquid can affect the rate at which the
excimers expand away from the track.  At high temperatures, the
excimers diffuse slowly because of scattering from phonons and rotons.
At low temperatures the excimers propagate ballistically into the cold
liquid in the absence of scattering. The excimer density drops much
more rapidly at low temperatures and as a consequence the rate of
generation of singlet species by the Penning process is strongly
temperature dependent.

The strong electric field dependence of the $1/t$ component of the
afterpulse results from the two separate ionization steps that are
involved in the production of these photons. Firstly, the initial
creation of triplet excimers by recombination is affected by the
electric field in the same manner as is the production of singlet
excimers responsible for the prompt signal. Secondly, the
recombination of the electron-ion pair produced by the Penning process
involving the triplet excimers is also influenced by the electric
field. Those recombinations that result in singlet species are
responsible for the $1/t$ component. Since the delayed generation of
the electron-ion pair occurs in isolation from other charges, the
recombination is geminate.

The Onsager theory of geminate recombination~\cite{ONS38} is not an
appropriate approach to calculating the effect of electric field on
the separation of charges created via Penning ionization in liquid
helium at low temperature. That theory assumes diffusion to be the
dominant process affecting the charge motion in the presence of the
field.

At low field and high temperatures where the ion mobilities are low
the motion of the charges is governed by viscous drag. In this case
the charges move along field lines, and whether two charges recombine
or not is dependent only on the magnitude and direction of their
initial separation with respect to an applied electric field.  If
$r_0$ the initial separation and $\theta_0$ is the angle between $r_0$
and the applied electric field $E$, the charges will recombine if
\begin{equation}
\label{eq:field}
 E<\frac{e}{4\pi \epsilon_0 r_0^2}\left ( 1+\tan^2(\theta_0/2)\right )\ . 
\end{equation}
Geminate recombination in LHe under this condition was studied in
Ref.~\cite{GUO11} using a $\beta$ source at 1.5~K.  However, at 1 K
where the mobility of ions is the order of $10^{-3}$~m$^2$/V$\cdot$s
The inertial term in the equation of motion becomes important at 
modest fields. In the limit of high mobility charges will recombine if
\begin{equation}
 E<\frac{e}{4\pi \epsilon_0 r_0^2}\ , 
\end{equation}
independent of orientation of the charges with respect to direction of
the field.  If this expression were used in conjunction with a
Gaussian distribution of charge separations to fit the field
dependence of the $1/t$ component of the afterpulse signal, the width
$b$ of the distribution would be unreasonably small, much less than
the 60~nm determined for the distribution of electrons about an alpha
track. Since the average energy of electrons created by the Penning
process responsible for the $1/t$ afterpulse component is comparable
to the average energy of secondary electrons from an alpha, the
inclusion of the inertial term in describing the ion motion does not
provide a satisfactory explanation to the field dependence.

At low temperatures and at zero pressure both positive ion snowballs
and electron bubbles moving in liquid helium create quantized vortex
rings~\cite{RAY64} when their velocities reach the order of 60
m/s. Since ion mobilities are the order of $10^{-3}$~m$^2/$V$\cdot$s,
at 1~K, fields of less than 1 kV/cm are sufficient for vortex
creation.  At low fields the charges remain attached to the rings that
they have produced, but in very high fields, greater than 25 kV/cm at
low temperatures, it has been found~\cite{NAN85} that electrons no
longer remain attached to vortices but create and shed vortex rings as
they move through the superfluid.

We have not attempted to apply the theory of vortex
creation by moving ions ~\cite{MUI84} to explain the field dependence of
the $1/t$ component of the afterpulses.  The energy a charge acquires
in moving in the field is transferred to the vortex, which increases
its diameter and decreases its velocity. Qualitatively, the vortices
introduce an enormous drag on the motion of the ions, which has the
effect of retarding their separation by the applied field. Presumably
vortices are the reason the field dependence of the afterpulses
produced by Penning ionization is not stronger than observed.

\subsubsection{Component independent of time}
The constant component of the afterpulses, independent of time, is most
naturally associated with triplet excimers whose density is constant and 
are uniformly distributed throughout the volume of the cell, the rate of 
destruction being balanced by the production from the 300~Bq $\alpha$ 
source. Destruction occurs in part by the spontaneous radiative decay 
of triplet excimers to the dissociated ground state and in part by the 
Penning process. 

The most notable features of the data seen in Fig.~\ref{fig:APconst}
are the apparent weak field dependence of this afterpulse component at
1~K and the substantial decrease in intensity as the temperature is
lower from 1~K to 0.4~K. These two properties are naturally explained
by the scintillation being principally due to the spontaneous
radiative decay of the excimers, the $a^3\Sigma_u^+ $ state having a
lifetime of 13~s in the bulk~\cite{MCK99}. While the electric field is
not expected to affect this process, it does lead to a decreased
recombination and density of triplets with increasing field. Below
about 0.6~K the mean free path of the excimers becomes long,
comparable to the dimensions of the helium cell, so that nonradiative
destruction of the excimers on solid surfaces decreases the photon
yield at low temperatures.

One observation remains unexplained.  Hereford and
coworkers~\cite{ROB73,ROB71} noted that the total scintillation below
0.5~K depended on the geometry of their cell, the smaller cell having
somewhat greater light output. The scintillation intensity decreased
with decreasing temperature to a minimum at 0.4~K and then increased
on going from 0.4~K to 0.2~K.  We have observed similar behavior of
the time-independent component of the afterpulse signal at low
temperatures (see Figs.~\ref{fig:APvsT} and \ref{fig:AtinvvsT}).

\subsection{Prediction of the prompt yield for LHe scintillation
  produced by the products of neutron capture on $^3$He}  
\subsubsection{Differences between scintillation induced by $\alpha$
  particles and scintillation induced by the products of the
  $^3$He($n$,$p$)$^3$H reaction} 
As discussed in the introduction, in the nEDM experiment, LHe
scintillation produced by the products of the neutron capture reaction
on $^3$He will be used as the neutron spin analyzer. In this reaction,
the $Q$ value of 760 keV is shared between the two reaction products,
a proton and a triton; the proton and the triton are emitted with
kinetic energies of 570 keV and 190 keV respectively. The range of a
570-keV proton in superfluid helium is 0.06~mm, corresponding to an
average energy deposition rate of $9.5\times
10^3$~eV/$\mu$m~\cite{NISTSTAR}. A 190-keV triton travels 0.018~mm in
superfluid helium depositing energy at a rate of $1.08\times
10^4$~eV/$\mu$m before it comes to rest~\cite{NISTSTAR}. In both
cases, the average energy deposition rate and hence the number of
ionizations per unit length along the track ($N_0$) are almost exactly
half of that for a 5.5-MeV $\alpha$ particle ($2.0\times
10^4$~eV/$\mu$m).

The radius of the ionization track $b$ is expected to be the same for
$\alpha$ particles and the products of neutron capture reaction on
$^3$He because the track radius is determined by the diffusion of the
electrons. Therefore, the ionization density inside the tracks created
by the reaction products of the $^3$He($n$,$p$)$^3$H reaction
is also almost exactly half of that in the track created by a 5.5-MeV
$\alpha$ particle.

The lower number of ionizations per unit length along the track (and
hence the lower ionization density) causes LHe scintillation produced
by the products of the $^3$He($n$,$p$)$^3$H reaction to differ from
$\alpha$-induced LHe scintillation in two aspects: 1) the lower
ionization density results in reduced quenching, thereby a larger
fraction of the deposited energy being emitted as prompt scintillation
compared to $\alpha$-induced scintillation. 2) As noted earlier in
Sec.~\ref{sec:promptdiscussion}, in the Kramers theory the fraction of
ions that recombine depends on a single parameter
$f=\sqrt{\pi}\epsilon_0 bE/(N_0e)$. Therefore, the lower number of
ionizations per unit length along the track means that the effect of
an electric field is larger.

\subsubsection{Model for quenching}
\label{sec:quenching}
The effect of the quenching by the nonradiative destruction of singlet
species by the Penning process can be described by the following equation:
\begin{eqnarray}
\label{eq:quench}
\frac{dn_s}{dt} & = & -\gamma_s (\kappa_{ss} n_s^2 + \kappa_{st} n_s n_t)
- \gamma_t \kappa_{tt} n_t^2 \nonumber \\ 
& & -  D_s\nabla^2n_s- \frac{n_s}{\tau_s},
\end{eqnarray}
where $n_s$ and $n_t$ are the densities of the singlet and triplet
species, respectively, $\gamma_s$ is the coefficient for bimolecular
decay involving singlet species, $\gamma_t$ is the coefficient for
bimolecular decay of the triplet species, $D_s$ is the diffusion
coefficient of the singlet species, and $\tau_s$ is the radiative
lifetime of the singlet species. $\kappa_{ss} = 7/4$, $\kappa_{st} =
2\times 3/4$, and $\kappa_{tt} = -1/7$ are numerical factors that come
from the fact that in the Penning process for every two excimers (or
excited atoms) destroyed, a new one is formed, one-quarter of the time
in the singlet state and three-quarters of the time in the triplet
state. 

The diffusion term can be neglected for the conditions we are
interested in, namely, the first $10^{-8}$~s of the hot track. 
Also the term describing the Penning ionization of two
triplet states can be neglected since $\gamma_s$ is sufficiently
larger than $\gamma_t$ (see below). 

Noting that $n_t$ changes with time more slowly than $n_s$, we solve
Eq.~(\ref{eq:quench}) for two limiting cases: 1) $n_t$ changes with
time keeping the $n_t/n_s$ ratio constant, that is $n_t = r_{ts} n_s$,
where $r_{ts}$ is a time independent constant. 2) $n_t$ does not vary
with time, that is $n_t = n_{t0}$, where $n_{t0}$ is the initial
triplet density. Solving Eq.~(\ref{eq:quench}) yields an analytical
solution of the same form for both cases. Integrating the last term of
the right hand side of Eq.~(\ref{eq:quench}) over time gives the
fraction, $f_s$, of the singlets that contribute to the prompt
scintillation escaping the bimolecular annihilation.
\begin{equation}
\label{eq:fs}
f_s = \frac{1}{n_{s0}}\int \frac{n_s}{\tau_s} dt = \frac{\log\left(1+\xi \right)}{\xi},
\end{equation}
where $n_{s0}$ is the initial singlet density and $\xi = \gamma_s'
\tau_s' n_{s0}$. For the case in which $n_t = r_{ts} n_s$, $\gamma_s'$
and $\tau_s'$ are given by $\gamma_s' = \gamma_s (\kappa_{ss} +
\kappa_{st} r_{ts})$ and $\tau_s' = \tau_s$. For the case where $n_t =
n_{t0}$, $\gamma_s'$ and $\tau_s'$ are given by $\gamma_s' = \gamma_s
\kappa_{ss}$ and $1/\tau_s' = 1/\tau_s + \gamma_s \kappa_{st} n_{t0}$.

For $\alpha$-induced LHe scintillation, the fraction of deposited
energy that is emitted as prompt scintillation was found to be
10\%~\cite{ADA95}. On the other hand, following a similar discussion
to the one in the beginning of Sec.~\ref{sec:promptdiscussion}, we
expect 23\% of the deposited energy to be emitted as prompt
scintillation in the absence of quenching, giving $f_s = 0.47$ for
$\alpha$-induced LHe scintillation. Solving Eq.~(\ref{eq:fs}) for $f_s
= 0.47$ gives $\xi = 3.5$ for $\alpha$-induced scintillation. The
numerical value of $\xi$ thus obtained allows us to make a crude
estimate of the coefficient for bimolecular decay of the singlet
species. For $n_t = r_{ts} n_s = 3n_s$, with $n_{s0} = 2.4\times
10^{16}$~cm$^{-3}$ and $\tau_s = 10^{-8}$~s, $\xi = 3.5$ gives
$\gamma_s = 2.3\times 10^{-9}$~cm$^3$/s for the singlet excimers. This
value for $\gamma_s$ is about an order of magnitude larger than the
value measured for the coefficient for bimolecular decay of the
triplet excimers~\cite{KET74a,ELT98,KAF00}. The difference could be
attributed to the suppression of this process for the triplets states
due to the electron spin flip required for the triplet excimer to go
to the dissociated ground state. Similar differences have been
observed in the deexcitation  rates of singlet and triplet states of
helium atoms when encountered by other atoms and compounds~\cite{YOS91}.

Assuming that the quenching fraction is the same for excimers and
excited atoms, setting $\xi=1.75 (=3.5/2)$ in Eq.~(\ref{eq:fs}) yields
$f_s = 0.58$ for LHe scintillation induced by the $^3$He($n$,$p$)$^3$H
reaction. It follows that the fraction of deposited energy emitted as
prompt scintillation is expected to be 13\% for LHe scintillation
produced by the products of the $^3$He($n$,$p$)$^3$H.

Note that this result is somewhat independent of the details of the
model. As long as the differential equation for $n_s$ has the form
$\dot{n_s} = -\gamma_s' n_s^2 - n_s/\tau_s'$, Eq.~(\ref{eq:fs}) holds
and we obtain the same value for $f_s$ for LHe scintillation induced
by the $^3$He($n$,$p$)$^3$H reaction.

\subsubsection{Prediction on the number of prompt EUV photons due to
  neutron capture on $^3$He} 
\begin{figure}
\includegraphics[width=9cm]{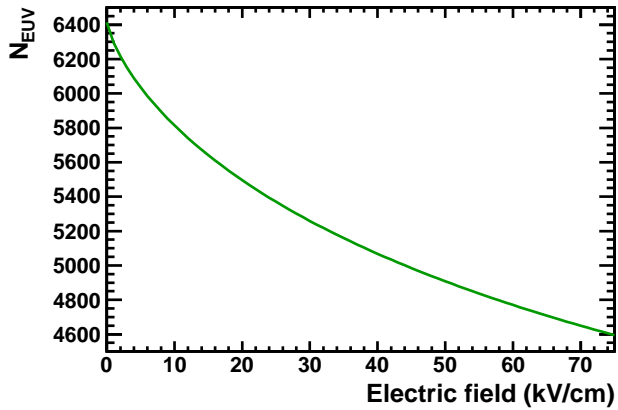}
\caption{(Color online) Predicted number of the prompt EUV photons for LHe
  scintillation produced by the products of the $^3$He($n$,$p$)$^3$H
  reaction with $x=0.65$ and $b=62$~nm.}
\label{fig:Neuv_3He_capture}
\end{figure}

With the model for quenching discussed above and Kramers' theory, we
can make a prediction on the number of the prompt EUV photons emitted
when a neutron is captured by a $^3$He atom in superfluid LHe. In
Fig.~\ref{fig:Neuv_3He_capture} the predicted number of the prompt EUV
photons for $^3$He($n$,$p$)$^3$H is plotted as a function of the
electric field.

Habicht~\cite{HAB98} in his thesis reports measurements of the
scintillation resulting from a number of different ionization sources
including the $^3$He($n$,$p$)$^3$H reaction in liquid helium at zero
field and 1.8~K. However, it is difficult to make a comparison with his results
given the lack of specificity of the experimental parameters such as
geometry, solid angle, etc. 

\section{Conclusion}

The prompt scintillation signal from $\alpha$ particles stopped in
helium exhibits a 15\% reduction in an electric field of
45~kV/cm. This field dependence is consistent with the current versus
field measurements of Gerritsen~\cite{GER48} and Kramers' analysis of
columnar recombination. We conclude using Kramers' theory that roughly
40\% of the scintillation results from species formed from atoms
originally promoted to excited states by the $\alpha$ particle and
60\% from excimers created by ionization and subsequent recombination,
with the electrons initially having a cylindrical Gaussian
distribution about the $\alpha$ track of 60~nm.

The delayed scintillation signal, the time dependence of which is
decomposed in the manner suggested by McKinsey {\it et
  al.}~\cite{MCK03}, exhibits stronger field and temperature
dependences than does the prompt scintillation. The stronger field
dependence is the consequence of the fact that the slow component of
afterpulses are from triplet excimers undergoing the Penning
ionization process followed by recombination forming singlet
excimers. As such, it receives the effect of the electric field twice,
once at the initial recombination producing triplet excimers, and once
at the recombination following the Penning process. The temperature
dependence involves the diffusion of the excimers away from the
$\alpha$ track into the surrounding bulk liquid.

\section{Acknowledgments}
The authors are grateful for the help and advice provided by the
following individuals at various stages of this work: E.~Bond,
M.~Fanning, S.~Currie, D.~G.~Haase, G.~Frossati, R.~Golub, W. Guo,
E.~Korobkina, H.-O.~Meyer, C.~Nelson, J.~Self, B.~Lozowski,
T.~Rinckel, P.~Childress, D.~Baxter, W.~M.~Snow, and A.~Edwards. This
work was supported by the US Department of Energy, the National
Science Foundation, and the DOE MIE No.71RE.

\end{document}